\documentclass[conference]{IEEEtran}
\IEEEoverridecommandlockouts
\usepackage{cite}
\usepackage{amsmath,amssymb,amsfonts}
\usepackage{algorithmic}
\usepackage{graphicx}
\usepackage{textcomp}
\usepackage{xcolor}

\usepackage{url}
\usepackage{cite}
\usepackage{tabularx}
\usepackage{booktabs}
\usepackage{CJKutf8}
\usepackage{hyperref}
\hypersetup{hidelinks}
\usepackage[caption=false,font=footnotesize]{subfig}
\usepackage[table]{xcolor}
\usepackage{multirow}
\usepackage{fontawesome5}

\def\BibTeX{{\rm B\kern-.05em{\sc i\kern-.025em b}\kern-.08em
    T\kern-.1667em\lower.7ex\hbox{E}\kern-.125emX}}
\begin{document}

\title{From Detection to Characterization: \\A Large-Scale Study of Ragebait on Japanese X
\thanks{This work was supported by JST ERATO (JPMJER2502).}
}

\author{
\IEEEauthorblockN{
Zhiyang Qi,
Kazuhiro Ito,
Jinghui Chen,
Hibiki Nakamura,
Zhangxuan Chen,\\
Erina Murata,
Masaki Chujyo,
and Fujio Toriumi
}
\IEEEauthorblockA{
\textit{Department of Systems Innovation, School of Engineering, The University of Tokyo}\\
7-3-1 Hongo, Bunkyo-ku, Tokyo 113-8656, Japan\\
\{zhiyangqi, kazuhiro-it, jinghuichen, harmonie-nkmr-00, chengmg, erinamurata, mchujyo\}@g.ecc.u-tokyo.ac.jp\\
tori@sys.t.u-tokyo.ac.jp
}
}

\maketitle

\begin{abstract}
Ragebait refers to online content intentionally designed to provoke anger or outrage and thereby increase attention and engagement. However, reliable large-scale detection and systematic analysis of ragebait remain limited, hindering efforts to understand its prevalence, impact, and mitigation. This study aims to develop an effective ragebait detection framework and to clarify the characteristics of ragebait at scale, providing a basis for understanding and mitigating emotionally provocative content online. We constructed a labeled dataset with the assistance of a large language model (LLM) and trained several Japanese language models for ragebait detection. The resulting ensemble classifier was then applied to a large-scale dataset of Japanese-language posts on X. Our analysis shows that ragebait is more prevalent in politically and socially contentious topics, including politics, discrimination, public health, and interpersonal conflict. Ragebait posts also spread faster and receive more negative reactions than non-ragebait posts, particularly anger, fear, disgust, sadness, and surprise. These findings demonstrate the utility of the proposed detector and provide a large-scale characterization of ragebait in Japanese online discourse.
\end{abstract}


\section{Introduction}
Ragebait refers to online content deliberately designed to provoke anger or outrage in order to attract attention and increase engagement. The growing prominence of this practice was reflected in the selection of \textit{rage bait} as the Oxford Word of the Year 2025, following a threefold increase in the use of the term over the preceding year \cite{oxford2025ragebait}. Unlike conventional clickbait, which often exploits curiosity gaps, ragebait captures attention by eliciting frustration, offense, or interpersonal conflict. Its increasing visibility highlights a broader shift in online attention strategies from stimulating curiosity to actively manipulating users' emotions \cite{potthast2018clickbait}.

A substantial body of research has demonstrated that anger and moral outrage play an important role in online engagement and information diffusion. Moral-emotional language has been associated with increased diffusion of political messages in social networks \cite{brady2017emotion}, while social feedback, such as likes and reposts, can reinforce users' subsequent expressions of outrage \cite{brady2021sociallearning}. Anger has also been found to spread efficiently across weak social ties \cite{fan2020weakties} and to generate larger and deeper information cascades than anxiety \cite{han2023anger}. These findings suggest that anger is not merely an individual response to online content: it can be amplified through social and platform-level mechanisms, potentially increasing polarization and degrading the quality of online discourse.

Nevertheless, research directly examining ragebait remains limited. Prior language-oriented work has discussed ragebait mainly in relation to intentionality and social-media discourse rather than as a scalable detection problem \cite{ohman2024textlength}. A recent study distinguished ragebait from information-oriented clickbait in news headlines and found that ragebait was associated with higher audience engagement \cite{shin2025emotion}. However, its focus was on professionally produced news headlines, leaving ragebait in general user-generated social media posts largely unexplored.

Detecting ragebait is also challenging because it cannot be reduced to negative sentiment, offensive language, or anger expression alone. A post may contain negative language without attempting to provoke its audience, while ragebait may use irony, exaggeration, selective framing, or deliberately unreasonable claims to trigger a reaction. Identifying ragebait therefore requires consideration of the communicative intent of the poster and the likely emotional response of readers. The lack of a scalable detection method makes it difficult to estimate how prevalent ragebait is, identify the users and topics associated with it, and examine how it spreads and influences audience reactions.



To address these limitations, we develop a ragebait detector for Japanese-language posts on X and conduct a large-scale analysis. Specifically, following the definition of ragebait, we first use a large language model (LLM) to determine whether individual posts constitute ragebait, thereby constructing a pseudo-labeled dataset for training text classification models. The resulting dataset contains 18,558 instances, with an equal number of ragebait and non-ragebait posts. We further ask two annotators to independently label a randomly sampled subset and confirm that the generated pseudo-labels exhibit a reasonable degree of reliability. Using this dataset, we then train text classifiers for ragebait detection. To improve robustness beyond any single model, we train multiple classifiers and combine the three best-performing models through majority voting. The resulting ensemble achieves an accuracy of 84.05\% and a Macro-F1 score of 84.04\%. 

We apply the ensemble classifier to posts extracted from a 1\% sample of X data collected between October 2022 and June 2023, comprising more than 150 million Japanese-language event records. In addition to quantifying the proportion of posts classified as ragebait within the analyzed data, we characterize the detected posts along multiple dimensions, including lexical characteristics, thematic content, diffusion dynamics, and the emotional reactions they receive.

Our analysis reveals several distinctive patterns. Ragebait posts are more prevalent in politically and socially contentious topics, including party politics, gender and discrimination, vaccines and infectious diseases, and moral conflict. They also tend to spread faster than non-ragebait posts. Moreover, replies and quote posts directed at ragebait contain substantially more negative emotions, particularly anger, fear, disgust, sadness, and surprise. Taken together, these findings provide converging evidence that the classifier captures a meaningful category of intentionally provocative content rather than merely negative language in general. Resources related to this study are publicly available\footnote{\faGithub\ \url{https://github.com/ZhiyangQi/japanese-x-ragebait}. To protect user privacy and avoid potential reputational harm, we do not release the original post text or the corresponding ragebait labels. Instead, we provide only the IDs of the source posts used for model training, the prompt used for ragebait labeling, and the models trained in this study.} for further research on ragebait detection and analysis.

\section{Related Work}

\subsection{Ragebait and Related Content}

Ragebait is related to clickbait, offensive language, hate speech, and toxic content. Prior work has developed datasets and detection methods for clickbait in news and social media \cite{biyani2016clickbait,chakraborty2016stop,potthast2018clickbait,indurthi2020clickbait}, as well as for offensive and abusive language \cite{davidson2017hate,waseem2017abuse,zampieri2019offenseval}. However, ragebait is defined not only by its linguistic form but also by the intention to provoke and the anticipated reaction of readers. Recent studies have begun to examine ragebait as a distinct phenomenon \cite{ohman2024textlength,shin2025emotion}, but large-scale detection and analysis of user-generated posts remain limited.

\subsection{Anger and Online Engagement}

Anger and moral outrage are strongly associated with online engagement and diffusion. Digital platforms can amplify moral outrage \cite{crockett2017outrage}, while moral-emotional language and social feedback increase its expression and diffusion \cite{brady2017emotion,brady2021sociallearning}. Anger also exhibits stronger social correlation and wider diffusion than other emotions \cite{fan2014anger,fan2020weakties,han2023anger}. Political out-group animosity and outrage-inducing misinformation are likewise associated with increased sharing \cite{rathje2021outgroup,mcloughlin2024misinformation}. These studies mainly examine emotions expressed in posts or their consequences, whereas our work focuses on detecting content intended to elicit such reactions.

\subsection{LLM-Assisted Annotation}

Recent advances in LLMs have enabled their use across a wide range of domain-specific NLP tasks \cite{qi2025kokorochat,qi2026less}. They have also increasingly been used for scalable text annotation \cite{gilardi2023chatgpt,pangakis2024supervised}. Classifiers trained on LLM-generated labels can approach the performance of models trained on human annotations \cite{pangakis2024supervised}, although subjective labels may be affected by annotation bias and model suggestions \cite{schroeder2025humanloop}. We therefore use an LLM to construct pseudo-labeled training data, validate a sampled subset through independent human annotation, and train separate encoder-based classifiers for large-scale ragebait detection.

\section{Dataset Construction}
\label{sec:dataset}

\subsection{Motivation and Overview}

\begin{figure*}[t]
\centering
\includegraphics[width=0.88\textwidth]{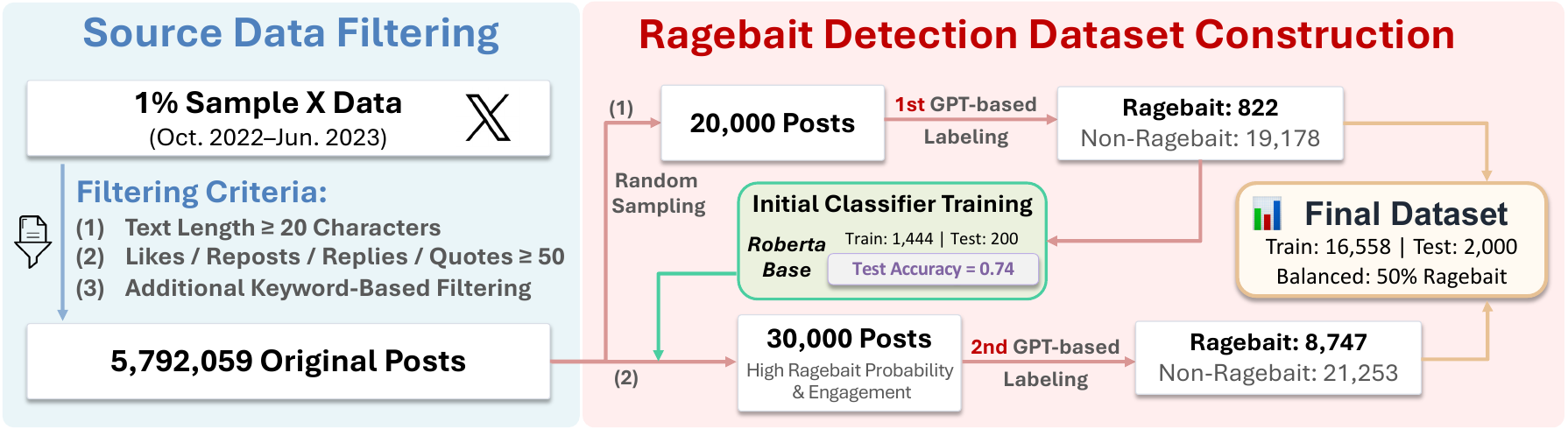}
\caption{Overview of the ragebait dataset construction pipeline.}
\label{fig:dataset_pipeline}
\end{figure*}

Constructing a ragebait dataset presents two major challenges. First, identifying ragebait requires assessing not only the surface wording of a post but also the author's likely intention to provoke and the emotional reactions that the post may elicit from readers, making large-scale manual annotation costly. Second, ragebait accounts for only a small proportion of randomly sampled social media posts, and simple random sampling therefore yields too few positive instances for training a reliable classifier. To address these challenges, we adopt a two-stage dataset construction procedure. We first use an LLM to assign pseudo-labels to a randomly sampled subset and train an initial ragebait detector. We then apply this detector to the remaining data to retrieve likely ragebait candidates, which are subsequently relabeled by the LLM. Fig.~\ref{fig:dataset_pipeline} provides an overview of the complete pipeline.

\subsection{Source Data and Filtering}

We use a 1\% sample of Japanese-language X data collected between October 2022 and June 2023. The raw data contain 153,849,869 event records, including original posts and engagement events such as reposts, replies, and quotes. Because each repost event preserves the corresponding original post together with its engagement counts at the time the event was recorded, we first extract the original posts associated with these events. We then retain posts containing at least 20 characters and require that at least one of the repost, like, reply, or quote counts exceeds 50. This filtering criterion allows us to focus on posts that received a certain level of public attention and for which engagement-based analyses are feasible. We further remove posts containing keywords associated with campaigns, lotteries, advertisements, promotions, and similar commercial content. After filtering, 5,792,059 original posts remain.

\subsection{First-Stage GPT-Based Labeling}

We randomly sample 20,000 posts from the filtered data and use GPT-5.4 mini\footnote{\url{https://developers.openai.com/api/docs/models/gpt-5.4-mini}} to determine whether each post constitutes ragebait. The labeling prompt is based on the definition of ragebait and asks the model to consider whether the post appears intended to anger, upset, or provoke readers and whether it is likely to elicit anger, indignation, disgust, or strong discomfort. GPT-5.4 mini assigns a binary label of \textsc{Yes} or \textsc{No} to each post. Among the 20,000 sampled posts, 822 are labeled as ragebait and 19,178 as non-ragebait. The resulting positive rate is only 4.11\%, illustrating that random sampling alone is inefficient for obtaining a sufficient number of ragebait examples.

To train an initial detector, we retain all 822 ragebait posts and randomly sample the same number of non-ragebait posts, resulting in a balanced seed dataset of 1,644 posts. We allocate 1,444 posts to the training set and 200 posts to the test set, with both sets containing equal numbers of ragebait and non-ragebait instances. We then fine-tune Rinna RoBERTa base\footnote{\url{https://huggingface.co/rinna/japanese-roberta-base}} as a binary classifier. The initial detector achieves an accuracy of 0.74 on the balanced test set. This detector is used only to retrieve additional candidate posts and is not used as the final classifier in our subsequent analyses.

\subsection{Detector-Assisted Data Expansion}

We apply the initial detector to the remaining filtered posts and obtain a predicted ragebait probability for each post. To efficiently collect additional positive examples, we select 30,000 posts with high predicted ragebait probabilities and high engagement counts. Incorporating engagement information enables us to prioritize posts that are both likely to be ragebait and sufficiently visible to generate observable diffusion and audience reactions.

The selected 30,000 candidates are then classified again by GPT-5.4 mini using the same definition and labeling criteria as in the first stage. In this second labeling stage, 8,747 posts are labeled as ragebait and 21,253 as non-ragebait. Compared with the initial random sample, the proportion of positive instances increases substantially, demonstrating that the initial detector is effective for retrieving likely ragebait candidates.

\subsection{Final Dataset}

We combine the labeled instances obtained from the two GPT-based labeling stages and remove duplicate posts, resulting in 9,279 unique posts labeled as ragebait. To construct a balanced dataset, we then randomly sample 9,279 posts from the available non-ragebait instances. The final dataset therefore contains 18,558 posts, consisting of 9,279 ragebait and 9,279 non-ragebait instances. We allocate 16,558 posts to the training set and 2,000 posts to the held-out test set. Both sets maintain a balanced class distribution, with ragebait and non-ragebait each accounting for 50\% of the instances. The held-out test set is used exclusively to evaluate the final ragebait classifiers described in Section~\ref{sec:detector}.

\subsection{Human Validation of GPT-Generated Labels}

Because ragebait judgments involve subjective assessments of communicative intention and anticipated emotional impact, we conduct an independent human evaluation of the GPT-generated labels. We randomly sample 200 posts from the pseudo-labeled dataset, including 100 posts labeled as ragebait and 100 labeled as non-ragebait. Two crowdworkers independently determine whether each post constitutes ragebait without access to the GPT-generated labels. The balanced sampling was intended to assess agreement across both label categories.

The crowdworkers are instructed to consider three criteria. First, they assess whether the post appears to have been written with the intention of angering, upsetting, or provoking readers. Second, they consider whether reading the post personally elicits anger, indignation, disgust, or strong discomfort. Third, they judge whether other readers would be likely to experience similar reactions. These criteria reflect both the presumed intention of the author and the anticipated emotional effects of the post.

\begin{table}[t]
\caption{Examples of ragebait and non-ragebait posts. For privacy protection, only English translations are shown; the translations were produced using GPT-5.5.}
\label{tab:ragebait_examples}
\centering

\scriptsize
\setlength{\tabcolsep}{4pt}
\renewcommand{\arraystretch}{1.08}

\begin{tabularx}{\columnwidth}{X}
\toprule

\textbf{Ragebait} \\
\midrule

\textit{(1) People who say that they are ``neither right-wing nor left-wing'' are almost always right-wing. I wonder why.} \\

\addlinespace

\textit{(2) When I see people who still believe government announcements and casually label opposing views as anti-vaccine, I used to get angry, but now I simply think that fools remain fools for life.} \\

\addlinespace

\textit{(3) There are almost no good, kind, righteous, loyal, principled, or decent people in this world. Everyone merely pretends to be decent, while many people are nothing but filth and scum.} \\

\midrule

\textbf{Non-Ragebait} \\
\midrule

\textit{(1) Izumi, leader of the Constitutional Democratic Party, announced that Hiroyuki Konishi would be removed from his position as the opposition's lead director on the Commission on the Constitution. Although it was described as a de facto dismissal, it was not an actual dismissal. The punishment was far too lenient.} \\

\addlinespace

\textit{(2) Saying that someone has a personality problem simply because they are experiencing mental distress will hurt them. Imagine how it would feel to already be suffering, to cry out without being understood, and instead to be treated as if you were mentally unstable. Such treatment can ultimately lead to discrimination and bullying.} \\

\addlinespace

\textit{(3) Hot Cook is trending again, but it is completely unsuitable for people who are not good at cleaning up, have a small kitchen sink, or cannot plan their meals several hours in advance. Please be careful. I paid nearly 50,000 yen for it but used it only during the first year. Even washing the inner pot is a chore.} \\

\bottomrule
\end{tabularx}

\end{table}

The agreement rate between Annotator A and GPT is 75.0\%, with Cohen's $\kappa=0.500$, while the agreement rate between Annotator B and GPT is 74.5\%, with $\kappa=0.498$. The two human annotators achieve an agreement rate of 78.5\%, with $\kappa=0.570$. This moderate level of inter-annotator agreement highlights the inherent subjectivity of ragebait judgments and suggests that perfect agreement is unlikely even among human annotators. Thus, the agreement between each human annotator and GPT is close to the agreement observed between the two human annotators. Although the GPT-generated labels should not be regarded as equivalent to manually established gold-standard labels, these results indicate that they possess a reasonable degree of reliability for constructing training data for ragebait detection. The classifier performance should therefore be interpreted in the context of this annotation ambiguity rather than against an assumption of perfectly consistent human judgments.

Table~\ref{tab:ragebait_examples} shows English translations of representative Japanese posts for which GPT and both human annotators assigned the same labels.

\section{Ragebait Detection Models}
\label{sec:detector}

\subsection{Experimental Setup}

Using the dataset described in Section~\ref{sec:dataset}, we train six Japanese pretrained language models for binary ragebait classification: Tohoku BERT base v3\footnote{\url{https://huggingface.co/tohoku-nlp/bert-base-japanese-v3}}, Rinna RoBERTa base, LINE DistilBERT base\footnote{\url{https://huggingface.co/line-corporation/line-distilbert-base-japanese}}, Waseda RoBERTa base\footnote{\url{https://huggingface.co/nlp-waseda/roberta-base-japanese}}, KU-NLP DeBERTa base\footnote{\url{https://huggingface.co/ku-nlp/deberta-v2-base-japanese}}, and KU-NLP DeBERTa large\footnote{\url{https://huggingface.co/ku-nlp/deberta-v2-large-japanese}}. The 16,558 training instances are further divided into training and validation subsets using stratified random sampling based on the Ragebait and Non-Ragebait labels, with 90\% used for training and 10\% for validation. The held-out test set of 2,000 instances is used only for final evaluation.

All models are fine-tuned for three epochs using AdamW. The training batch size is set to 16, while the validation batch size is set to 32. We search learning rates from $\{1\times10^{-5}, 2\times10^{-5}, 3\times10^{-5}\}$ and select the checkpoint with the highest Macro-F1 score on the validation set. We report Accuracy, Precision, Recall, and Macro-F1 on the held-out test set.

\subsection{Classification Results}

Table~\ref{tab:classification_results} presents the classification performance of the six individual models. Among the single models, Tohoku BERT base v3 achieves the highest Accuracy and Macro-F1, reaching 83.40\% and 83.38\%, respectively. It also obtains the highest Recall of 87.00\%. Rinna RoBERTa base achieves the highest Precision of 83.10\%, while LINE DistilBERT base also shows competitive overall performance.

\begin{table*}[t]
\caption{Performance of the ragebait detection models on the test set. The best result in each column is shown in bold.}
\label{tab:classification_results}
\centering
\footnotesize
\setlength{\tabcolsep}{5pt}
\renewcommand{\arraystretch}{1.0}

\begin{tabular}{lcccc}
\toprule
\textbf{Model}
& \textbf{Accuracy (\%)}
& \textbf{Precision (\%)}
& \textbf{Recall (\%)}
& \textbf{Macro-F1 (\%)} \\
\midrule

Tohoku BERT base v3
& 83.40
& 81.16
& \textbf{87.00}
& 83.38 \\

Rinna RoBERTa base
& 82.90
& \textbf{83.10}
& 82.60
& 82.90 \\

LINE DistilBERT base
& 82.65
& 81.79
& 84.00
& 82.65 \\

Waseda RoBERTa base
& 80.65
& 77.99
& 85.40
& 80.61 \\

KU-NLP DeBERTa large
& 80.60
& 81.29
& 79.50
& 80.60 \\

KU-NLP DeBERTa base
& 79.70
& 77.15
& 84.40
& 79.66 \\

\midrule

LINE + Rinna + Tohoku Majority Vote
& \textbf{84.05}
& 82.65
& 86.20
& \textbf{84.04} \\

\bottomrule
\end{tabular}
\end{table*}

\subsection{Ensemble Classifier}

Although the three best-performing individual models achieve similar overall performance, they are based on different architectures and pretraining resources and may therefore exhibit different error patterns. Relying on a single model could make the final predictions more sensitive to model-specific biases and occasional misclassifications. To improve robustness and reduce the influence of any one model, we combine the binary predictions of Tohoku BERT base v3, Rinna RoBERTa base, and LINE DistilBERT base using majority voting.

The resulting ensemble achieves an Accuracy of 84.05\% and a Macro-F1 score of 84.04\%, outperforming all individual models on both metrics. Its Precision and Recall are 82.65\% and 86.20\%, respectively, indicating a favorable balance between detecting ragebait posts and limiting false positives. Based on its stronger overall performance and improved robustness, we adopt the majority-voting ensemble as the final ragebait detector for the large-scale analyses described in the following sections.

\section{Large-Scale Analysis of Posts on X}
\label{sec:large_scale_analysis}

We apply the ensemble detector described in Section~\ref{sec:detector} to a large-scale dataset derived from a 1\% sample of Japanese-language X collected between October 2022 and June 2023. From 153,849,869 event records, we extract and deduplicate the original posts associated with repost events, because these records preserve the engagement counts of the original posts at the time each repost was observed, enabling subsequent analyses of diffusion. To focus the analysis on general public discourse and reduce contamination from commercial or promotional content, we filter out posts matching a predefined list of keywords related to advertisements, lotteries, campaigns, and promotions. This results in 17,435,881 posts, of which 1,118,931 posts (6.42\%) are classified as ragebait and 16,316,950 as non-ragebait. We use this data to analyze lexical characteristics, thematic patterns, and diffusion speed.

\subsection{Lexical Characteristics}
\label{sec:lexical_characteristics}

\begin{figure}[t]
\centering
\subfloat[Ragebait.\label{fig:wordcloud_ragebait}]{%
    \includegraphics[width=0.7\columnwidth]{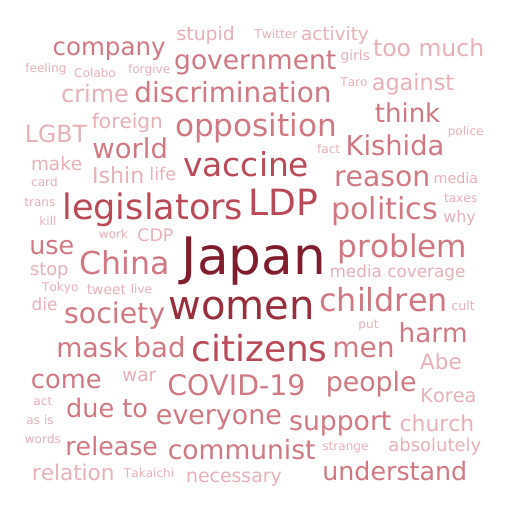}%
}

\vspace{2mm}

\subfloat[Non-ragebait.\label{fig:wordcloud_nonragebait}]{%
    \includegraphics[width=0.7\columnwidth]{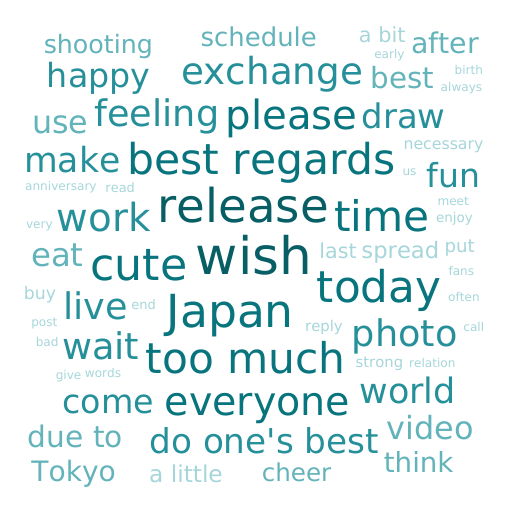}%
}

\caption{Word clouds of ragebait and non-ragebait posts. Word size represents within-group frequency. English translations are from GPT-5.5.}
\label{fig:wordclouds}
\end{figure}

\begin{figure*}[t]
\centering
\includegraphics[width=0.70\textwidth]{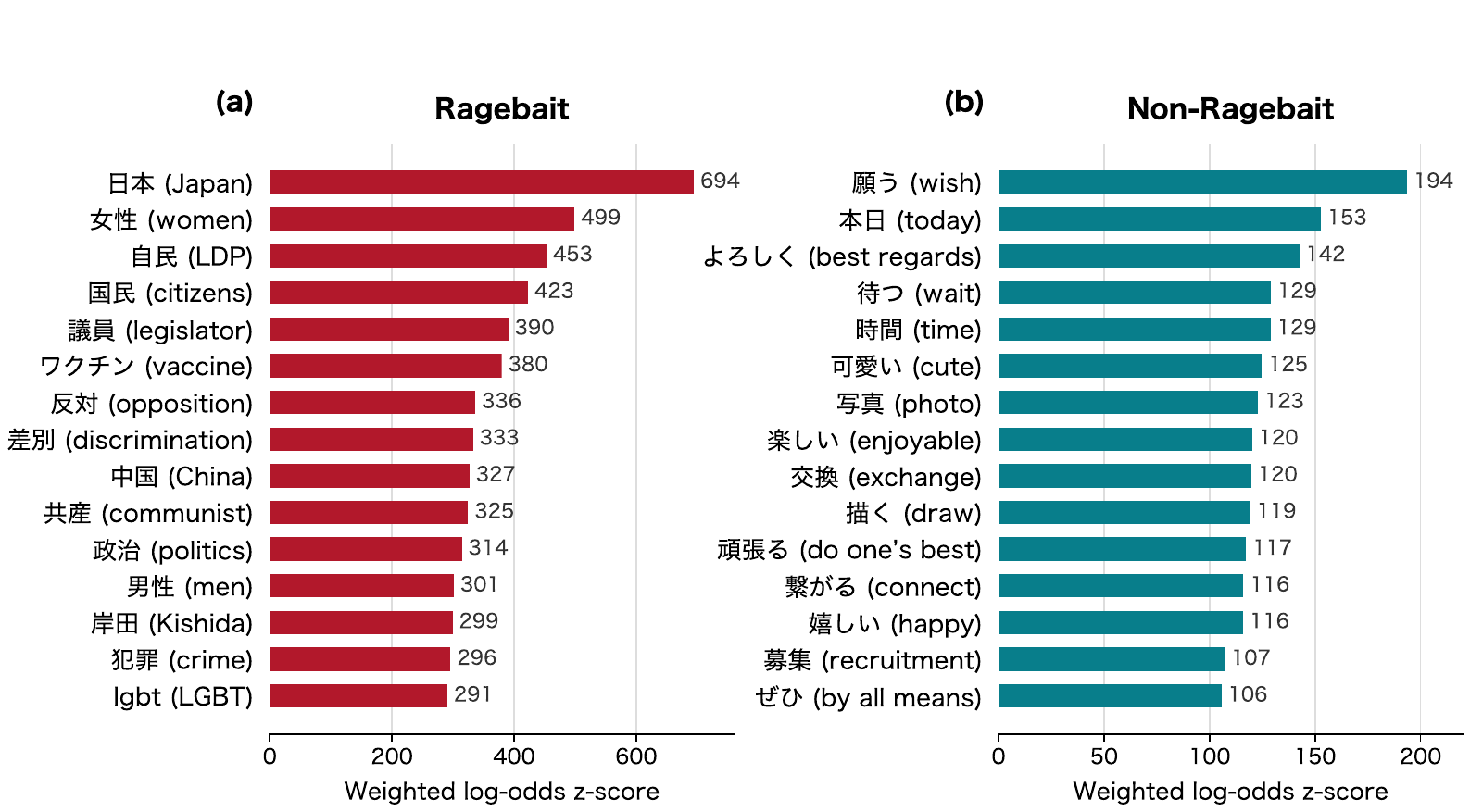}
\caption{Top 15 distinctive words for ragebait and non-ragebait posts ranked by weighted log-odds $z$-score. }
\label{fig:weighted_log_odds}
\end{figure*}

We first examine whether ragebait and non-ragebait posts exhibit distinct lexical patterns. This analysis provides an interpretable overview of the language associated with each group and helps assess whether the detector identifies semantically coherent differences rather than merely general negative sentiment.

We preprocess the post text and perform morphological analysis using \texttt{fugashi}\footnote{\url{https://github.com/polm/fugashi}}. We retain nouns, verbs, adjectives, and adverbs, while removing highly general terms. We then calculate token frequencies separately for ragebait and non-ragebait posts and construct the word clouds shown in Fig.~\ref{fig:wordclouds}. Larger words indicate higher frequencies within each group.
Considering that raw frequency can be dominated by terms that are common to both groups, we further compute weighted log-odds $z$-scores \cite{monroe2008fightinwords}. This measure identifies terms that are overrepresented in one group relative to the other while accounting for differences in data size and overall token frequency. Fig.~\ref{fig:weighted_log_odds} presents the 15 most distinctive terms for each group.

Taken together, the word clouds and weighted log-odds results reveal clear lexical differences between the two groups. Ragebait posts more frequently contain terms related to politics, social identity, public health, and social conflict, including ``Japan,'' ``women,'' ``LDP,'' ``vaccine,'' ``discrimination,'' ``China,'' ``crime,'' and ``LGBT.'' By contrast, non-ragebait posts more frequently contain terms associated with everyday communication, creative activities, and recreational content, such as ``wish,'' ``today,'' ``best regards,'' ``cute,'' ``photo,'' ``enjoyable,'' ``draw,'' and ``happy.'' These patterns indicate that, within the analyzed data, posts classified as ragebait are more lexically concentrated around contentious public issues, whereas non-ragebait posts more often concern everyday interaction and leisure-related topics.

\subsection{Differences in Topic Prevalence}
\label{sec:topic_prevalence}

Lexical analysis captures individual words, but it does not directly reveal broader thematic structures. We therefore analyze the thematic differences between ragebait and non-ragebait posts using Structural Topic Modeling (STM) \cite{roberts2019stm}. STM estimates latent topics in a document collection and can incorporate document-level metadata, enabling us to compare topic prevalence between the two groups while considering temporal variation.

For this analysis, we use 1,073,827 ragebait posts and 13,028,190 non-ragebait posts after additional preprocessing for topic modeling. Because the two groups differ greatly in size, we sample 100,000 posts from each group for STM training while preserving the monthly distribution of the original data. We use the ragebait label and posting month as document metadata. The ragebait label is used to compare topic prevalence between the two groups, while posting month is included to reduce the influence of temporal shifts in topic composition.

We compare models with $K \in \{15,20,25\}$ topics and select the final number of topics based on held-out likelihood, exclusivity, and semantic coherence. Based on these criteria, we adopt $K=25$ and apply the trained STM to the full posts.

\begin{figure}[t]
\centering
\includegraphics[width=\columnwidth]{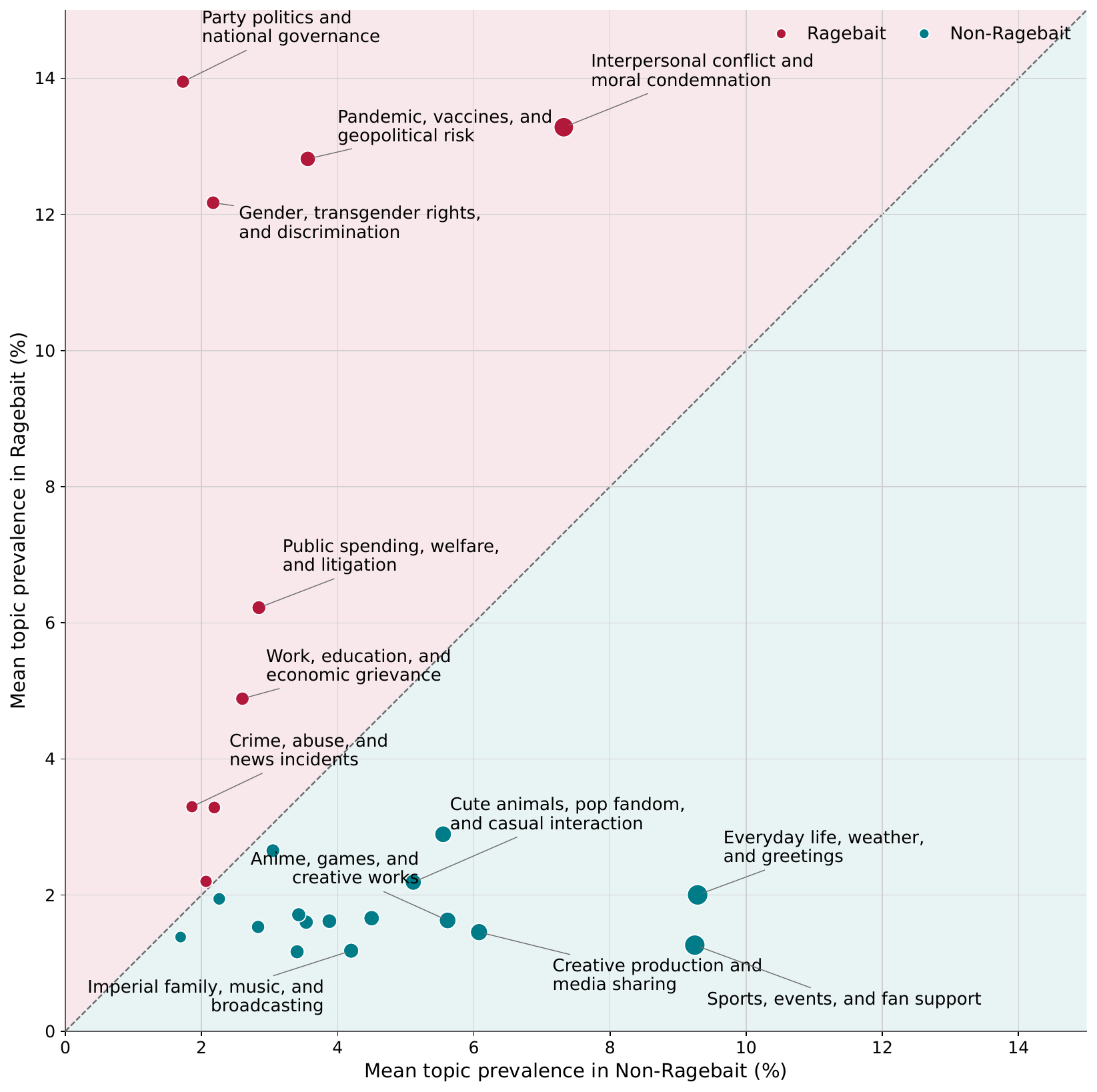}
\caption{Mean topic prevalence in ragebait and non-ragebait posts. Points above the diagonal are more prevalent in ragebait posts, while points below it are more prevalent in non-ragebait posts.}
\label{fig:stm_topic}
\end{figure}

Fig.~\ref{fig:stm_topic} shows that ragebait posts have higher topic prevalence for party politics and national governance, gender and discrimination, vaccines and infectious diseases, interpersonal conflict and moral condemnation, and public spending, welfare, and litigation. In contrast, non-ragebait posts show higher prevalence for topics related to sports, daily interaction, creative production, anime and games, and information exchange.

The four most prevalent ragebait-associated topics---party politics and national governance, interpersonal conflict and moral condemnation, vaccines and infectious diseases, and gender and discrimination---account for 52.22\% of the mean topic prevalence in ragebait posts, compared with 14.78\% for the same topics in non-ragebait posts. This result indicates that ragebait is not evenly distributed across topics, but is concentrated in a limited set of socially contentious themes where disagreement and emotional reactions are more likely to occur.

\subsection{Diffusion Dynamics}
\label{sec:diffusion_dynamics}

We next compare the diffusion dynamics of ragebait and non-ragebait posts. This analysis is motivated by the possibility that ragebait may not only differ in content, but also in how quickly it elicits engagement after being posted.

To enable temporal analysis, we restrict the data to original posts that were observed at least five times in the repost-event records. This allows us to track changes in engagement counts across multiple time points for the same post. We examine four engagement indicators: reposts, likes, replies, and quotes.

For each post, we sort its observations chronologically and compute the increase in each engagement count between two consecutive observations. We then divide this increment by the elapsed time between the observations to obtain an hourly growth rate. To reduce the influence of account size, we normalize the growth rate by the author's follower count and report the median growth per 10,000 followers. Finally, we group the results by elapsed time since posting and compare ragebait and non-ragebait posts within each time bin.

\begin{table*}[t!]
\caption{Binary sentiment and fine-grained emotion occurrence rates in replies and quote posts. All columns except $N$ are percentages; shaded cells indicate the higher value within each comparison.}
\label{tab:emotion_reactions}
\centering
\footnotesize
\renewcommand{\arraystretch}{1.12}

\resizebox{0.98\textwidth}{!}{
\begin{tabular}{llrrrrrrrrrrr}
\toprule
\multirow{2}{*}{\textbf{Reaction}}
& \multirow{2}{*}{\textbf{Target}}
& \multirow{2}{*}{\textbf{N}}
& \multicolumn{2}{c}{\textbf{Binary Sentiment}}
& \multicolumn{8}{c}{\textbf{Fine-Grained Emotions}} \\
\cmidrule(lr){4-5}
\cmidrule(lr){6-13}
&
&
& \textbf{Positive}
& \textbf{Negative}
& \textbf{Joy}
& \textbf{Trust}
& \textbf{Ant.}
& \textbf{Sur.}
& \textbf{Sad.}
& \textbf{Fear}
& \textbf{Ang.}
& \textbf{Disg.} \\
\midrule

\multirow{2}{*}{Replies}
& Ragebait
& 819,406
& 47.98
& \cellcolor{gray!20}\textbf{52.02}
& 48.62
& 19.41
& 55.91
& \cellcolor{gray!20}\textbf{64.08}
& \cellcolor{gray!20}\textbf{42.47}
& \cellcolor{gray!20}\textbf{59.93}
& \cellcolor{gray!20}\textbf{26.58}
& \cellcolor{gray!20}\textbf{57.18} \\

& Non-ragebait
& 33,256,332
& \cellcolor{gray!20}\textbf{78.56}
& 21.44
& \cellcolor{gray!20}\textbf{81.64}
& \cellcolor{gray!20}\textbf{34.78}
& \cellcolor{gray!20}\textbf{70.58}
& 50.53
& 25.52
& 29.33
& 4.27
& 19.41 \\

\midrule

\multirow{2}{*}{Quotes}
& Ragebait
& 215,287
& 36.51
& \cellcolor{gray!20}\textbf{63.49}
& 43.12
& 22.28
& 61.37
& \cellcolor{gray!20}\textbf{75.17}
& \cellcolor{gray!20}\textbf{42.04}
& \cellcolor{gray!20}\textbf{64.00}
& \cellcolor{gray!20}\textbf{35.53}
& \cellcolor{gray!20}\textbf{69.46} \\

& Non-ragebait
& 3,605,330
& \cellcolor{gray!20}\textbf{77.26}
& 22.74
& \cellcolor{gray!20}\textbf{82.72}
& \cellcolor{gray!20}\textbf{38.19}
& \cellcolor{gray!20}\textbf{79.02}
& 62.85
& 21.47
& 24.97
& 6.40
& 21.18 \\

\bottomrule
\end{tabular}
}
\end{table*}

\begin{figure}[t]
\centering
\includegraphics[width=\columnwidth]{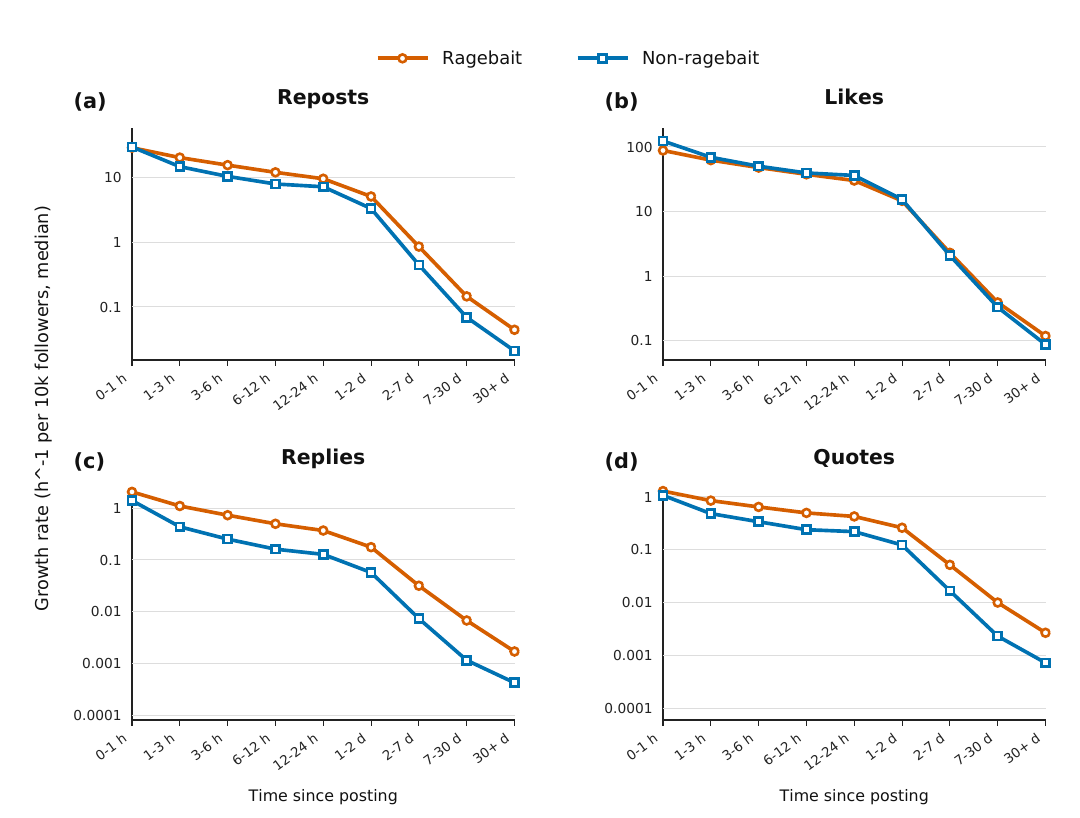}
\caption{Median engagement growth per 10,000 followers over time for ragebait and non-ragebait posts: (a) reposts, (b) likes, (c) replies, and (d) quotes.}
\label{fig:diffusion_speed}
\end{figure}

Fig.~\ref{fig:diffusion_speed} shows that the growth rate is highest immediately after posting for all four engagement indicators and gradually declines over time. For reposts, ragebait posts show slightly higher growth than non-ragebait posts after the first hour and continue to maintain an advantage across several later time bins. For likes, the two groups follow very similar trajectories, with only small differences throughout the observed period.

More pronounced differences appear for replies and quotes. In both cases, ragebait posts show consistently higher growth rates than non-ragebait posts, and this gap remains visible from the first few hours after posting to several days later. 
These results suggest that ragebait posts are not only reposted at a slightly faster rate, but more importantly, they are more likely to trigger replies and quotes, which often reflect discussion, disagreement, or reactive engagement. In this sense, ragebait appears to facilitate interaction patterns that extend beyond passive exposure and are more likely to sustain public response over time.

\subsection{Emotional Reactions in Replies and Quotes}
\label{sec:emotional_reactions}

Finally, we examine the emotional characteristics of replies and quote posts directed at ragebait and non-ragebait posts. Unlike the preceding analyses, which focus on the detected original posts themselves, this analysis uses a separate reaction dataset consisting of original posts that received at least one reply or quote. After deduplication, we obtain 34,523,063 original posts. We classify these original posts using the three-model majority-voting detector, resulting in 846,962 ragebait posts (2.45\%)\footnote{The 2.45\% classifier-positive rate differs from the 6.42\% reported in the preceding analyses because the two datasets use different inclusion criteria: the former includes source posts receiving at least one reply or quote, whereas the latter is derived from source posts recovered from repost-event records.} and 33,676,101 non-ragebait posts (97.55\%). Based on this classification, we compare the sentiment and fine-grained emotions expressed in replies and quote posts.

\subsubsection{Binary Sentiment}

We first conduct binary sentiment classification to examine whether reactions to ragebait are more negative than reactions to non-ragebait. We use a Japanese sentiment classification model\footnote{\url{https://huggingface.co/llm-book/bert-base-japanese-v3-wrime-sentiment}} trained on WRIME \cite{kajiwara2021wrime} and classify each reply and quote post as either positive or negative.
Table~\ref{tab:emotion_reactions} shows that reactions to ragebait are substantially more negative than reactions to non-ragebait. Among replies to ragebait posts, 52.02\% are classified as negative, compared with 21.44\% for replies to non-ragebait posts. The difference is even larger for quote posts: 63.49\% of quotes directed at ragebait are negative, compared with 22.74\% for quotes directed at non-ragebait. This suggests that ragebait posts are more likely to elicit negative audience reactions, particularly through quote posts.

\subsubsection{Fine-Grained Emotion Analysis}

We further analyze fine-grained emotions using a Japanese emotion classification model fine-tuned on WRIME\footnote{\url{https://huggingface.co/Munek/bert-large-japanese-v2-finetuned-wrime}}. We examine eight basic emotions: joy, trust, anticipation, surprise, sadness, fear, anger, and disgust. Because the model treats each emotion independently, a single reply or quote post may be assigned multiple emotions. Following the model setting, we apply ROC-optimized thresholds for each emotion and regard an emotion as present when its predicted score exceeds the corresponding threshold.

Table~\ref{tab:emotion_reactions} shows that replies and quote posts directed at ragebait contain higher rates of sadness, surprise, anger, fear, and disgust than those directed at non-ragebait. The contrast is especially clear for anger and disgust. For example, anger appears in 26.58\% of replies and 35.53\% of quotes directed at ragebait, whereas it appears in only 4.27\% of replies and 6.40\% of quotes directed at non-ragebait. Similarly, disgust appears in 57.18\% of replies and 69.46\% of quotes directed at ragebait, compared with 19.41\% and 21.18\% for non-ragebait.

By contrast, joy, anticipation, and trust are more frequent in reactions to non-ragebait posts. These results indicate that ragebait does not merely increase negative sentiment in general; rather, it is associated with specific negative emotions such as anger, fear, and disgust. The tendency is particularly pronounced in quote posts, suggesting that ragebait is more likely to trigger reactive or critical forms of engagement. As our operational definition of ragebait explicitly considers its potential to elicit anger, indignation, disgust, or strong discomfort, these observed reaction patterns provide converging evidence consistent with the definition used in this study.

\section{Conclusion}
\label{sec:conclusion}

This study developed a ragebait detector for Japanese-language posts on X and conducted a large-scale analysis of ragebait. We constructed an LLM-generated pseudo-labeled dataset of approximately 18K posts, validated the annotation reliability through independent human evaluation, and trained an ensemble classifier that achieved 84.05\% accuracy and a Macro-F1 score of 84.04\%.
Applying the detector to large-scale X data, we found that only a small proportion of the analyzed posts were classified as ragebait. Posts classified as ragebait were disproportionately associated with politically and socially contentious topics and showed stronger growth in replies and quotes. Reactions to these posts also contained more negative sentiment and higher rates of anger, fear, and disgust than reactions to posts classified as non-ragebait.

Overall, these findings suggest that ragebait is a distinct form of emotionally provocative content rather than merely negative or offensive language. Although relatively uncommon, it appears to have a disproportionate capacity to stimulate conflict-oriented interaction and negative emotional responses. The proposed detector provides a practical basis for large-scale analysis and future mitigation efforts.



\section*{Acknowledgment}

This work was supported by JST ERATO (JPMJER2502).

\vspace{12pt}

\end{document}